\documentclass[aps,twocolumn,prd,showpacs,showkeys,preprintnumbers,superscriptaddress,nobibnotes,floatfix,longbibliography,nofootinbib]{revtex4-1}

\usepackage[T1]{fontenc} 
\usepackage[compat=1.1.0]{tikz-feynman}
\usepackage{pifont}
\usepackage[mathscr]{eucal}
\usepackage{multirow}
\usepackage{graphicx}
\usepackage{amsfonts}
\usepackage{amsmath}
\usepackage{amssymb}
\usepackage{epsfig}
\usepackage{slashed}
\usepackage{dcolumn}%
\usepackage{ulem}
\usepackage{color}
\usepackage{setspace}
\usepackage{tikz}
\usepackage{cancel}
\usepackage{tabularx}
\definecolor{myred}{RGB}{255, 0, 0}
\definecolor{myblue}{RGB}{0, 0, 255}
\definecolor{mygreen}{RGB}{0, 128, 0}

\usepackage[colorlinks,
linkcolor=blue,
citecolor=blue,
urlcolor=blue
]{hyperref}
\begin{document}
\title{The phenomenon of the axion kinetic misalignment with a generic PQ-breaking operator}

\author{Xiangwei Yin}
\email{yinxiangwei@cqu.edu.cn}
\affiliation{Department of Physics and Chongqing Key Laboratory for Strongly Coupled Physics, Chongqing University, Chongqing 401331, P.R. China}

\author{Ligong Bian}
\email{lgbycl@cqu.edu.cn}
\affiliation{Department of Physics and Chongqing Key Laboratory for Strongly Coupled Physics, Chongqing University, Chongqing 401331, P.R. China}

\begin{abstract}
We investigate the phenomenology induced by generic PQ-breaking operators within the axion kinetic misalignment framework. We analyze their impact on the relic density of axion dark matter (DM), the PQ quality problem, axion-mediated fifth-force, as well as Big Bang Nucleosynthesis (BBN) and Cosmic Microwave Background (CMB) constraints. A nonzero initial axion velocity gives rise to brief periods of early matter domination and axion kinetic domination, leading to a nonstandard cosmological evolution. We compute the resulting gravitational wave (GW) signal from global cosmic strings and find that, because these nonstandard epochs are extremely short, the signal is highly suppressed and beyond the reach of existing experiments. Finally, we perform a parameter space scan, identify the regions and benchmark point that are consistent with all experimental constraints.
\end{abstract}
\maketitle

\newpage
\section{Introduction}
The $U(1)_A$ problem~\cite{Weinberg:1975ui} arises because the expected light pseudoscalar associated with $U(1)_A$ does not appear in the hadron spectrum. This issue was resolved by recognizing that $U(1)_A$ is not a true symmetry in quantum chromodynamics (QCD), as revealed by the non-trivial topological QCD vacuum~\cite{Callan:1976je,Jackiw:1976pf}. This topological structure allows a CP-violating term in the QCD Lagrangian, which is tightly constrained by the neutron electric dipole moment (nEDM) and requires $\bar{\theta} < 10^{-10}$. The extreme smallness of $\bar{\theta}$ gives rise to the strong CP problem. Peccei and Quinn (PQ) introduced a new global $U(1)_{\mathrm{PQ}}$ symmetry that is spontaneously broken at a high scale~\cite{Peccei:1977hh,Peccei:1977ur,Weinberg:1977ma,Wilczek:1977pj}.
The associated pseudo Nambu Goldstone boson, the axion, dynamically relaxes the effective $\bar{\theta}$ parameter to zero, thereby eliminating the CP violation.
Axion is a well-motivated candidate of cold dark matter (DM), and is conventionally assumed to be produced through the misalignment mechanism (See Ref.~\cite{DiLuzio:2020wdo,Hook:2018dlk} for the review and the cosmological phenomenology can be found in Refs.~\cite{Marsh:2015xka,OHare:2024nmr}). In this scenario, the axion field begins its evolution near a static initial value, and the relic density is determined by the initial misalignment angle and the axion mass. An axion decay constant of order $10^{11}~\mathrm{GeV}$, corresponding to a mass of roughly $50~\mu\mathrm{eV}$, yields the saturated relic density.

In recent years, the kinetic misalignment mechanism~\cite{Co:2019jts,Chang:2019tvx} has attracted attention as an extension of the traditional misalignment mechanism. Its central feature is that the axion field is allowed to possess a nonzero initial velocity ($\dot{\theta}\neq 0$), which substantially delays the onset of oscillations. As a consequence, one can obtain the correct relic density even for significantly heavier axions (about $0.1\sim 100~\mathrm{meV}$) or for smaller decay constants ($10^{8} \sim 10^{11}~\mathrm{GeV}$), thereby relaxing the stringent constraints on axion decay constant or mass. Such a nonzero initial velocity can arise naturally from explicit PQ symmetry breaking in the early Universe and is closely analogous to the rotational dynamics of a complex scalar field in the Affleck-Dine mechanism~\cite{Affleck:1984fy,Dine:1995kz}.

In this work, we investigate the phenomenology induced by generic PQ-breaking higher-dimensional operators within the framework of the axion kinetic misalignment mechanism. We analyze their impact on the relic density of axion DM, which is determined by the initial axion velocity induced by explicit PQ breaking and the subsequent dynamics of the radial mode. We further consider the PQ quality problem, fifth-force constraints, and bounds from Big Bang Nucleosynthesis (BBN), Cosmic Microwave Background (CMB).
The combined motion of the axion and the radial mode traces an elliptic trajectory in field space, during which the energy density redshifts as $a^{-3}$, resulting in a matter-dominated era.
As the radial mode relaxes toward its vacuum, only the kinetic energy of the axion remains, leading to a subsequent kination-dominated era. 
We study the resulting gravitational wave (GW) signal from global cosmic strings~\cite{Gouttenoire:2019kij,Chang:2019mza,Co:2021lkc,Gouttenoire:2021jhk,Eroncel:2024rpe}. The inserted eras enhance the spectrum during matter domination and suppress it during kination. However, both epochs are exceedingly short, yielding a GW signal that is highly suppressed and effectively undetectable.
The exceptionally weak GW signal originates from the fact that, at the Lagrangian level, it is not possible to simultaneously obtain a sizable GW amplitude and maintain the correct cosmological evolution. This is different from Ref.~\cite{Co:2021lkc,Gouttenoire:2021jhk}, where the analysis is carried out in terms of intermediate variables.
Finally, we perform a parameter-space scan of the model and identify the viable region and benchmark point consistent with all relevant experimental constraints.

This paper is organized as follows. In Sec.~\ref{Sec2}, we review the kinetic misalignment mechanism. Sec.~\ref{Sec3} discusses the phenomenological implications, including the axion relic density, the PQ quality problem, fifth-force constraints, and BBN and CMB constraints. In Sec.~\ref{Sec4}, we discuss the GW signal from global cosmic strings, and scan the parameter space and identify benchmark point compatible with experimental limits. We conclude in Sec.~\ref{Sec5}.

\section{Kinetic misalignment}\label{Sec2}
In this section, we briefly review the kinetic misalignment mechanism~\cite{Co:2019jts,Chang:2019tvx} for axion DM production.

We begin with the PQ complex scalar field
\begin{equation}
    \Phi \equiv \frac{1}{\sqrt{2}}\left(S+f_a\right) e^{i \frac{a}{f_a}},
\end{equation}
where $S$ is the radial mode, $a=f_a \theta$ is the angular mode (the axion field), and $f_a$ is the decay constant. 
In the kinetic misalignment mechanism, the axion field starts with a nonzero velocity $\dot{\theta}\neq 0$ rather than a static displacement. This motion can be described by a rotation of $\Phi$, and can be quantified by the PQ charge density and the corresponding yield
\begin{equation}
    n_\theta \equiv \dot{\theta} S^2~,\quad Y_\theta \equiv \frac{n_\theta}{s}~,
\end{equation}
where $s$ the entropy density. $Y_\theta$ remains constant as long as PQ charge and entropy are conserved. 
At early times, the axion carries large kinetic energy $K=\dot{\theta}^2 f_a^2 / 2$ and freely overshoots the periodic potential barriers, so it is not initially trapped. As the universe expands, it falls below the barrier height $V_{\max }=2 m_a^2(T) f_a^2$. At this point, when $\dot{\theta} = 2 m_a$, the axion becomes trapped in a potential minimum and begins coherent oscillations. Because this trapping occurs later than the conventional onset temperature, the start of oscillations is delayed. The axion relic density is 
\begin{equation}
\Omega_a h^2 = \frac{\rho_{a_0}}{\rho_{\mathrm{crit}}}h^2=\frac{2m_a Y_{\theta} s_0}{\rho_{\mathrm{crit}}}h^2 \simeq0.12\left(\frac{10^9 \mathrm{GeV}}{f_a}\right)\left(\frac{Y_\theta}{38}\right)~,
\label{II-3}
\end{equation}
where we use the axion yield $Y_a=2 Y_\theta$ (the factor 2 arises because anharmonic effects near the top of the axion potential~\cite{Co:2019jts}), $\rho_{\mathrm{crit}}\simeq 1.05 \times 10^{-5}h^2 \mathrm{GeV} \mathrm{cm}^{-3}$, $s_0\simeq 2891\mathrm{cm}^{-3}$, and the zero-temperature axion mass $m_a \simeq 5.7\times 10^{-12}\mathrm{GeV}\left(10^{9} \mathrm{GeV} / f_a\right)$.

The origin of the nonzero initial velocity of the axion can be attributed to higher-dimensional operators that explicitly break the PQ symmetry, generating a potential gradient along the angular direction. This initial kick dynamics of the PQ field set the value of $Y_\theta$, which directly controls the axion relic density. We consider the following general potential for the complex scalar field $\Phi$
\begin{equation}
\begin{aligned}
    V(\Phi)&=V_0(\Phi)+V_{\cancel{\mathrm{PQ}}}(\Phi)\\
    &=\lambda\left(|\Phi|^2-\frac{f_a^2}{2}\right)^2+ \frac{\lambda_n|\Phi|^{2 m}\left(e^{-i \delta_n} \Phi^n+e^{i \delta_n} \Phi^{\dagger^n}\right)}{M_{\mathrm{Pl}}^{2m+n-4}}.
\end{aligned}
\end{equation}
In the presence of explicit PQ breaking, the operator dimension $2m+n$ must be at least greater than 7, provided no fine-tuning of the phase $\delta_n$. The second term breaks the PQ symmetry explicitly, together with the dynamics of the radial mode, modifies the relic density of axion DM.
Moreover, this operator spoils the axion solution to the strong CP problem and also induces axion-mediated CP-even forces.
The non-standard cosmological evolution associated with the rotating field is subject to bounds from BBN and the CMB.

\section{Experimental constraints}\label{Sec3}
In this section, we discuss the experimental constraints on the kinetic misalignment mechanism from various observations, including the relic density of axion DM, PQ quality problem, fifth-force, and BBN and CMB.
\subsection{Axion DM}
In this subsection, we analyze the axion DM relic density in the presence of higher-dimensional operators that break the PQ symmetry.

\begin{figure}[h]
    \centering
    \includegraphics[width=0.48\textwidth]{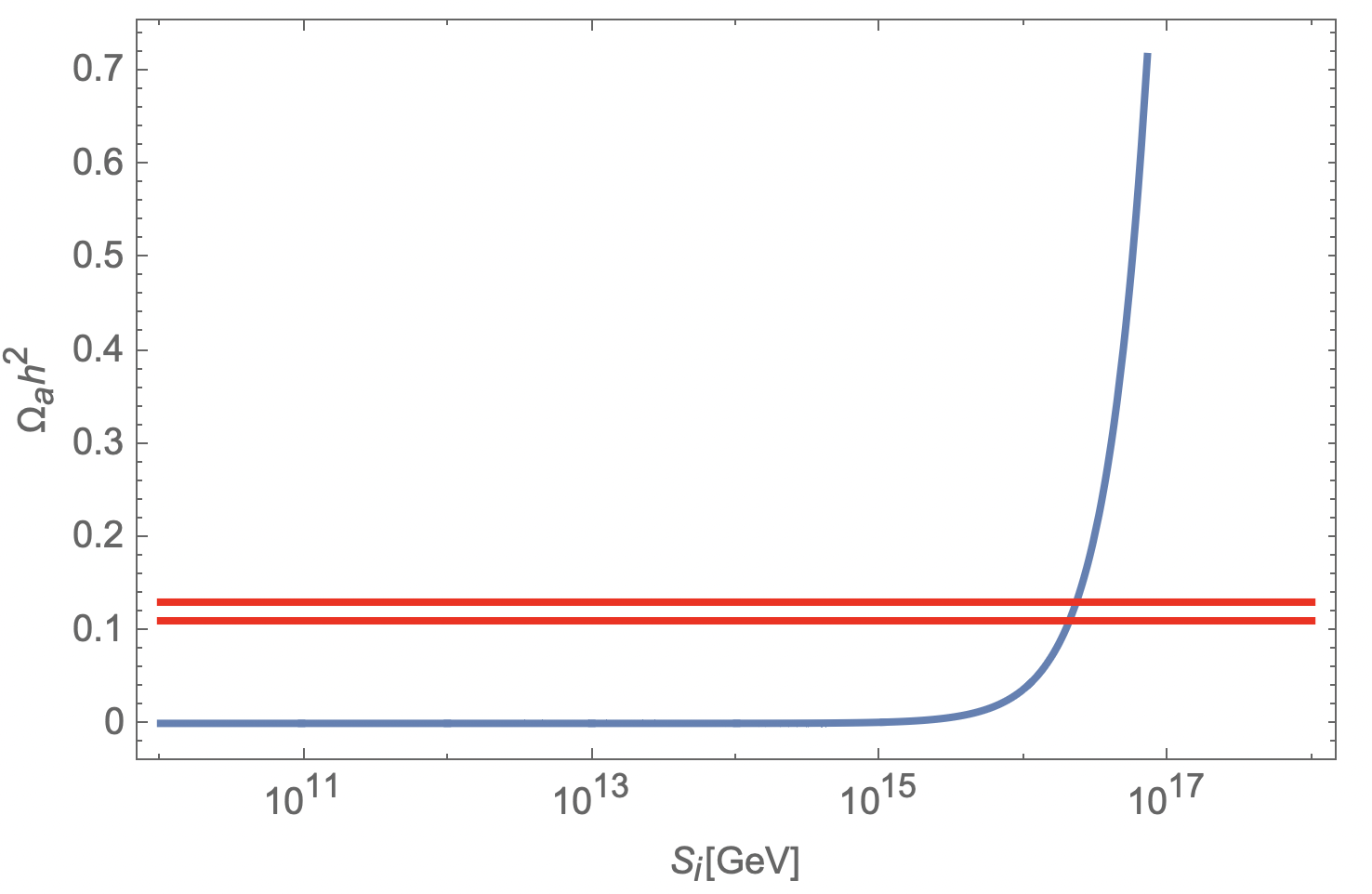}
    \caption{Axion relic density as a function of the initial field value. The red line is the saturated relic density. We employ $m=5$, $n=1$, $\lambda_n=10^{-10}$, $\lambda=10^{-20}$, $\delta_n=0$, $\epsilon=1$, and $f_a=10^{8}~\mathrm{GeV}$.}
    \label{fig:axion_relic_density}
\end{figure}

The evolution of scalar field in the PQ breaking potential can be written  as
\begin{equation}
    \ddot{\Phi}+3 H \dot{\Phi}+\frac{\partial}{\partial \Phi}\left(V_0(\Phi)+V_{\cancel{\mathrm{PQ}}}(\Phi)\right)=0.
\end{equation}
This leads to coupled equations of motion for the radial and angular components. The higher-dimensional operators generate a potential gradient along the angular direction of $\Phi$, thereby inducing angular motion. The dynamics of the radial mode influence the relic density of axion DM, with its mass determined by
\begin{equation}
\begin{aligned}
m_S^2 &= 3\lambda S^2 + 2^{2-m-\frac{n}{2}}\lambda_n  \cos\delta_n  \left(2m+n-1\right)\left(m+\tfrac{n}{2}\right) \\
      &\quad \times M_\mathrm{Pl}^{4-2m-n}\, S^{2m+n-2}.
\end{aligned}
\end{equation}
The radial mode begins to oscillate when the mass exceeds $3H$, with $H=1.66g_{\star}^{1/2}\frac{T^2}{M_{\mathrm{Pl}}}$. The  oscillation temperature is
\begin{equation}
    T_{\mathrm{osc}}=\left(\frac{m_S M_{\mathrm{Pl}}}{1.66 \times 3 g_{\star}^{1/2}}\right)^{1/2}.
\end{equation}
We define the charge density of radial mode and the ratio as
\begin{equation}
    n_s=\frac{V\left(\Phi\right)}{m_S\left(\Phi\right)},\quad \epsilon= \frac{n_\theta}{n_s},
\end{equation}
where $\epsilon$ denotes the ratio of the angular to radial potential gradient. It quantifies how closely the angular motion follows a circular trajectory. In particular, $\epsilon=1$ corresponds to a perfectly circular orbit.  
The relic density of axion DM is influenced by the dynamics of the radial mode, which affects the angular motion and consequently the axion relic density. The yield is 
\begin{equation}
    Y_\theta=\frac{n_\theta}{s(T_{\mathrm{osc}})},
\end{equation}
which is a redshift-invariant quantity and is determined by the initial conditions. Then we can use Eq.~\eqref{II-3} to calculate the axion relic density. In FIG.~\ref{fig:axion_relic_density}, we illustrate the dependence of the axion relic density on the initial field value $S_i$. The red line represents the saturated relic density, indicating that the allowed values of $S_i$ are highly constrained.

\subsection{The PQ quality and fifth-force}
In this subsection, we examine the constraints on the PQ quality and axion mediated fifth-force arising from higher-dimensional operators that break the PQ symmetry explicitly~\cite{Kamionkowski:1992mf}.

\begin{figure}[!htp]
    \centering
    \includegraphics[width=0.48\textwidth]{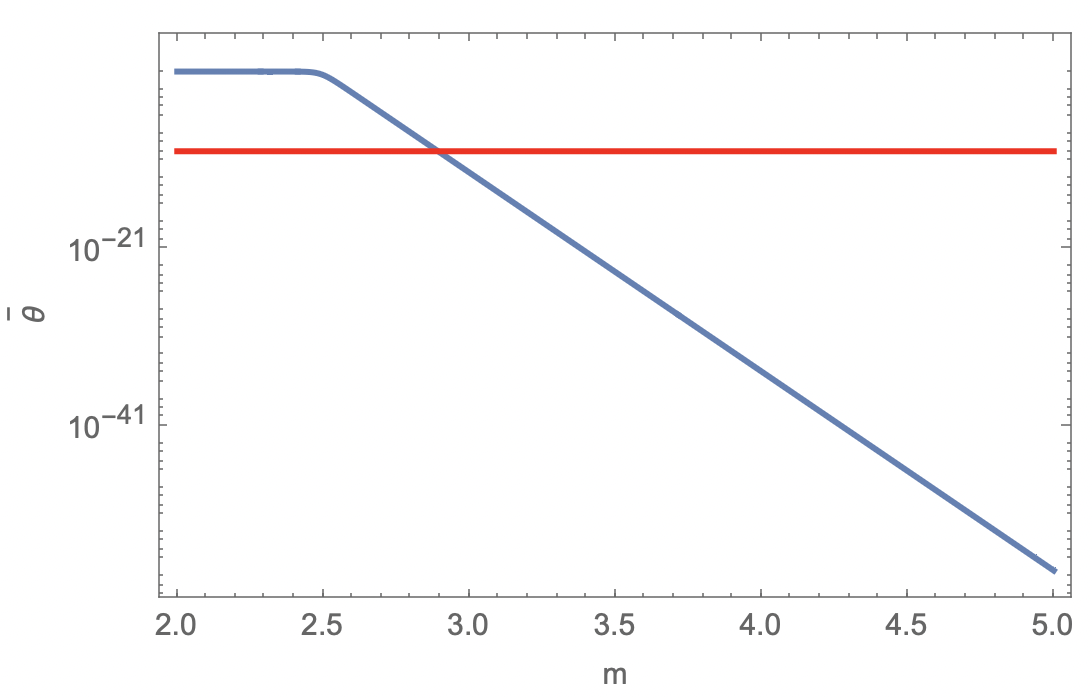}
    \caption{Constraints on the PQ quality from nEDM measurements. The shaded region is excluded. We set $n=1, \delta_n=0.1, \lambda_n=10^{-20}, f_a=10^{8} \mathrm{GeV}$.}
    \label{fig:PQ_quality}
\end{figure}

The PQ quality problem arises because the PQ mechanism relies on an almost exact global $U(1)_{PQ}$ symmetry to solve the strong CP problem. In quantum field theory, however, global symmetries are not fundamental and can be violated by Planck-suppressed higher-dimensional operators. Such terms can explicitly break $U(1)_{PQ}$, generating an additional contribution to the axion potential that shifts its minimum away from the CP-conserving point. A generic form of the potential including both the QCD contribution and the PQ breaking term can be written as
\begin{equation}
    V(a)=f_a^2\left[m_{\cancel{\mathrm{PQ}}}^2\mathrm{cos}(\frac{na}{f_a}-\delta_n)+m_a^2\mathrm{cos}\frac{a}{f_a}\right],
\end{equation}
where $m_{\cancel{\mathrm{PQ}}}^2=\lambda_n M_{\mathrm{Pl}}^2(\frac{f_a}{\sqrt{2}M_{\mathrm{Pl}}})^{2m+n-2}$ encodes the effect of the PQ breaking effect, $m_a$ is the QCD induced axion mass. The corresponding $\bar{\theta}$ term is modified to
\begin{equation}
    \overline{\theta}=\frac{n~\text{sin}\delta_n}{n^2~\text{cos}\delta_n+R},
\end{equation}
where $R\equiv m_a^2/m^2_{\cancel{\mathrm{PQ}}}$. The experimental upper limit on the nEDM constrains the effective $\bar{\theta}$ parameter to be less than $10^{-10}$~\cite{Abel:2020pzs}. FIG.~\ref{fig:PQ_quality} illustrates the dependence of $\bar{\theta}$ on the operator dimension $m$, and the red line is nEDM bound. 

In the standard PQ mechanism, the axion dynamically cancels the $\theta$ term, ensuring that the effective CP-violating angle $\theta_{\mathrm{eff}}$ relaxes to zero. This guarantees the absence of observable CP violation in the strong sector and, consequently, forbids any CP-even scalar coupling of the axion to matter. However, when the PQ symmetry is explicitly broken by higher-dimensional operators, the cancellation is no longer exact. The axion vacuum is shifted away from the CP-conserving point, resulting in a residual effective angle $\theta_{\mathrm{eff}}\neq 0$. This induces a CP-violating scalar interaction between the axion and nucleons
\begin{equation}
    \mathcal{L}_a^{\mathrm{CPV}}=-g_{a N}^S a \bar{N} N~,
\end{equation}
where the induced scalar coupling is approximately~\cite{Moody:1984ba,Bertolini:2020hjc,DiLuzio:2025zma}
\begin{equation}
    \small
    g_{a N}^S \simeq \frac{\theta_{\mathrm{eff}}}{f_a} \frac{m_u m_d}{m_u+m_d} \frac{\langle N| \bar{u} u+\bar{d} d|N\rangle}{2} \simeq 2 \cdot 10^{-11} \theta_{\mathrm{eff}}\left(\frac{10^{9} \mathrm{GeV}}{f_a}\right) .
\end{equation}
\begin{figure}
    \centering
    \includegraphics[width=0.45\textwidth]{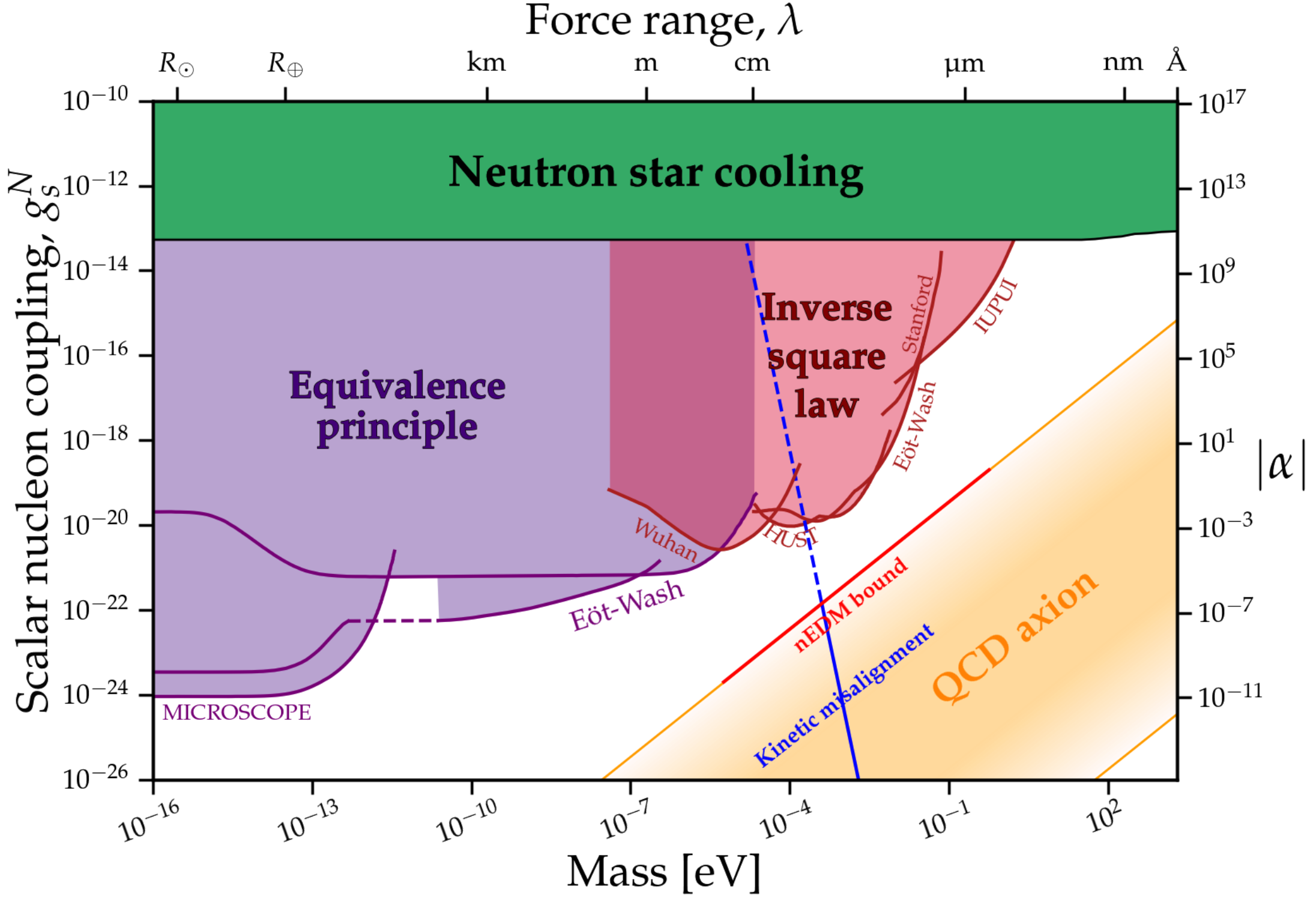}
    \caption{The red line shows the nEDM bound. The solid blue line indicates the fifth-force constraint on $g_{aN}$ in the kinetic misalignment scenario, while the blue dashed line is excluded by the nEDM bound. Evidently, for QCD axions the nEDM bound is stronger than the fifth-force limit. we employe $\delta=\frac{\pi}{2}, m=4, n=1, \lambda_n=10^{-10}$. Figure from Ref.~\cite{AxionLimit}.}
\label{fig:fifth_force}
\end{figure}
In combination with the usual pseudoscalar axion coupling, this interaction allows the axion to mediate macroscopic fifth-forces, including monopole-monopole and monopole-dipole interactions, which are testable in equivalence-principle and inverse-square-law experiments. As shown in FIG~\ref{fig:fifth_force}, the solid blue line corresponds to the parameter space of the axion-nucleon coupling in the kinetic misalignment scenario, obtained by varying $f_a$ while keeping the other parameters fixed, whereas the red line indicates the nEDM bound satisfied by the QCD axion. The blue dashed line is excluded by the nEDM bound. It is therefore evident that the constraint from fifth-force experiments is weaker than that implied by PQ quality. Although the solid blue line remains within the QCD axion parameter space, this is a consequence of the unmodified $m_a$-$f_a$ relation. The kinetic misalignment does not enlarge the intrinsic parameter space of the QCD axion but instead modifies the region in which the axion can account for the observed DM.


\subsection{BBN and CMB}
In this subsection, we investigate the constraints from BBN and CMB on the kinetic misalignment mechanism.
This mechanism induces a period of early matter-dominated era, after which the universe enters a kination-dominated era. The specific epoch at which this occurs determines whether it affects BBN and the CMB~\cite{Co:2021lkc}.

\begin{figure}
    \centering
    \includegraphics[width=0.45\textwidth]{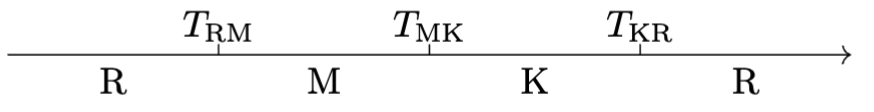}
    \caption{Cosmological history including an axion-induced kination era.}
    \label{fig:kination}
\end{figure}

In axion cosmology, a kination-dominated era can naturally emerge within the kinetic misalignment mechanism. The process begins when the PQ complex scalar field is displaced far from its minimum in the early universe. Higher-dimensional PQ breaking operators provide a kick to the angular direction, setting the field into rotation. Initially, the motion is a mixture of radial and angular dynamics, corresponding to an early matter-dominated era with energy density scaling as $\rho \propto a^{-3}$.
As the radial oscillations are damped (through interactions or cosmic expansion), the system evolves toward nearly circular motion. Once the radial field relaxes to its vacuum $f_a$, the axion energy is almost purely kinetic. At this point the universe enters a kination era, where the axion kinetic energy dominates and the energy density redshifts as $\rho \propto a^{-6}$. The FIG.~\ref{fig:kination} illustrates this cosmological history, showing the successive eras and the associated critical temperatures.

The energy density of the axion field is given by $\rho_{\theta}=\dot{\theta}^2 S^2$. The standard radiation energy density is expressed as
$\rho_{R}=\frac{\pi^2}{30}g_*T^4$. The transition tremperature is determined by equating these two densities, leading to the following critical temperatures
\begin{equation}
    T_{\mathrm{RM}}=\frac{4}{3}\frac{g_{*s}}{g_*}\dot{\theta}Y_{\theta}.
    \label{Sec3_C_TRM}
\end{equation}

The entropy density of the axion is given by $s=\frac{2\pi^2}{45}g_{*s} T^3$. Once the axion field relaxes to its vacuum $f_a$, the system enters a kination-dominated era, during which the entropy density can also be expressed as $s=\tfrac{n_\theta}{Y_\theta}=\tfrac{\dot{\theta} f_a^2}{Y_\theta}$. The transition temperature of the matter-kination epoch is
\begin{equation}
    T_{\mathrm{MK}}=\left(\frac{45}{2\pi^2g_{*s}}\frac{\dot{\theta}f_a^2}{Y_{\theta}}\right)^{1/3}.
    \label{Sec3_C_TMK}
\end{equation}

During the kination era, the axion energy density is dominated by its kinetic component, $\rho_a = \frac{1}{2}\dot{\theta}^2 f_a^2 \propto a^{-6}$. As the universe expands, this energy density redshifts faster than radiation, and the kination epoch ends once $\rho_a$ falls below the radiation energy density $\rho_R$. The transition temperature can be written as
\begin{equation}
    T_{\mathrm{KR}}=\frac{3\sqrt{15}}{2\pi}(\frac{g_*^{1/2}}{g_{*s}})\frac{f_a}{Y_{\theta}}.
    \label{Sec3_C_TKR}
\end{equation}

Theses three characteristic temperatures satisfy a strict hierarchy dictated by cosmic cooling, namely
$T_{\rm RM} > T_{\rm MK} > T_{\rm KR}$. Consequently, to evade cosmological bounds from BBN and the CMB, it is sufficient to require that the lowest of these, $T_{\rm KR}$, lies above the critical temperature set by observations. In practice, ensuring $T_{\rm KR} \gtrsim {\cal O}({\rm MeV})$ is sufficient to avoid both BBN and CMB constraints~\cite{Co:2021lkc}.

\section{GWs}\label{Sec4}
In this section, we discuss the GWs signal generated by global cosmic strings.
Cosmic strings  can form during a $U(1)$ symmetry-breaking phase transition in the early universe. Once formed, the string network evolves toward a scaling regime in which its correlation length grows linearly with cosmic time~\cite{Klaer:2017qhr,Gorghetto:2018myk,Vaquero:2018tib,Klaer:2019fxc,Gorghetto:2020qws}. 
During the fragmentation of the string network, closed loops are continuously generated and typically formed with size $\alpha t_i$ at cosmic time $t_i$, with a formation rate
\begin{equation}
\frac{d n_{\mathrm{loop}}}{d t_i} \simeq F_\alpha \frac{C_{\mathrm{eff}}\left(t_i\right)}{\alpha} t_i^{-4} ,
\end{equation}
where $F_\alpha\sim 0.1$ and  $C_{\mathrm{eff}}\sim{\cal O}(1)$ denotes loop emission parameter.
These loops undergo relativistic oscillations and continuously lose their energy through two main channels: gravitational radiation and Goldstone boson emission. 
The continuous energy loss modifies the loop size after its formation. A loop born at time $t_i$ with initial length $\alpha t_i$ evolves according to
\begin{equation}
l(\tilde{t})=\alpha t_i-(\Gamma G \mu+\kappa)\left(\tilde{t}-t_i\right),
\end{equation}
where $\Gamma G \mu$ and $\kappa$ are the gravitational radiation and Goldstone boson emission rates, respectively. The loops radiate GW with emission frequencies $f_{\mathrm{emit}} \simeq 2k/l$, where $k$ is a positive integer.
The loop radiates energy and shrinks until its length has decreased to half of its initial value. This moment defines the time of maximal GW emission, $\tilde{t}_M$, and corresponds to a loop size $l_M=\alpha t_i/2$. At this stage the fundamental emission mode $k=1$ has frequency $f_M=2/l_M$, which after redshift gives the observed frequency today
\begin{equation}
f=\frac{4}{\alpha t_i}\left[\frac{a\left(\tilde{t}_M\right)}{a\left(t_0\right)}\right]~,
\label{Sec3_D_1}
\end{equation}
where $t_{i}$ is the formation time, $\tilde{t}_M$ is emission time, and $t_0$ is the present time. For global string, $t_i \simeq \tilde{t}_M$. Considering that GWs are emitted at the critical temperatures $T_{\mathrm{KR}}$, $T_{\mathrm{MK}}$ and $T_{\mathrm{RM}}$, and using Eq.~(\ref{Sec3_D_1}), the peak frequencies are 
\begin{equation}
    \small
\begin{aligned}
    f_{\mathrm{KR}}&=3.89\times 10^{-8} \mathrm{Hz}~g_{*}^{1/2}\left(T_{\mathrm{KR}}\right) \left(\frac{0.1}{\alpha}\right) \left(\frac{T_{\mathrm{KR}}}{\mathrm{GeV}}\right) \left(\frac{g_{*s}(T_0)}{g_{*s}(T_{KR})}\right)^{1/3},\\
    f_{\mathrm{MK}}&=f_{\mathrm{KR}}\left(\frac{g_*(T_{\mathrm{MK}})}{g_*(T_{\mathrm{KR}})}\right)^{1/3}\left(\frac{T_{\mathrm{MK}}}{T_{\mathrm{KR}}}\right)^{4/3},\\
    f_{\mathrm{RM}}&=f_{\mathrm{MK}}\left(\frac{g_*(T_{\mathrm{RM}})}{g_*(T_{\mathrm{MK}})}\right)^{1/6}\left(\frac{T_{\mathrm{RM}}}{T_{\mathrm{MK}}}\right)^{2/3},\\
\end{aligned}    
\end{equation}
and the peak amplitude at the end of kination is~\cite{Gouttenoire:2019kij,Gouttenoire:2021jhk}
\begin{equation}
    \small
\begin{aligned}
    \Omega_{\mathrm{GW}, \mathrm{KR}} &\simeq\left(1.2 \times 10^{-18}\right)\left(\frac{f_a}{10^{15} \mathrm{GeV}}\right)^4 \exp \left(2 N_{\mathrm{KR}}\right)\\
    &\log ^3\left[\left(2.2 \times 10^{18}\right)\left(\frac{f_a}{10^{15} \mathrm{GeV}}\right)\left(\frac{\alpha}{0.1}\right)^2\left(\frac{10^9 \mathrm{GeV}}{E_{\mathrm{KR}}}\right)^2\right]~,
    \label{Sec3_D_GW}
\end{aligned}
\end{equation}
where $N_{\mathrm{KR}}=\frac{1}{6}\log(\frac{T_{\mathrm{MK}}}{T_{\mathrm{KR}}})$, $E_{\mathrm{KR}}=\rho_{\mathrm{KR}}^{1/4}$ is the energy scale at which the kination ends, and the scaling behavior of GWs spectrum is
\begin{equation}
    \Omega_{\mathrm{GW}}(f)=\Omega_{\mathrm{GW},\mathrm{KR}} 
    \begin{cases}
        1 &  f < f_{\mathrm{KR}}~, \\ 
        \frac{f}{f_{\mathrm{KR}}} &  f_{\mathrm{KR}} \leq f \leq f_{\mathrm{MK}}~, \\ 
        \frac{f_{\mathrm{MK}}}{f_{\mathrm{KR}}}\left(\frac{f_{\mathrm{MK}}}{f}\right)^{1/3} &  f_{\mathrm{MK}} \leq f \leq f_{\mathrm{RM}}~, \\ 
        \frac{f_{\mathrm{MK}}}{f_{\mathrm{KR}}}\left(\frac{f_{\mathrm{MK}}}{f_{\mathrm{RM}}}\right)^{1/3} & f> f_{\mathrm{RM}}~.
    \end{cases}
\end{equation}

\begin{figure}[!htp]
    \centering
    \includegraphics[width=0.4\textwidth]{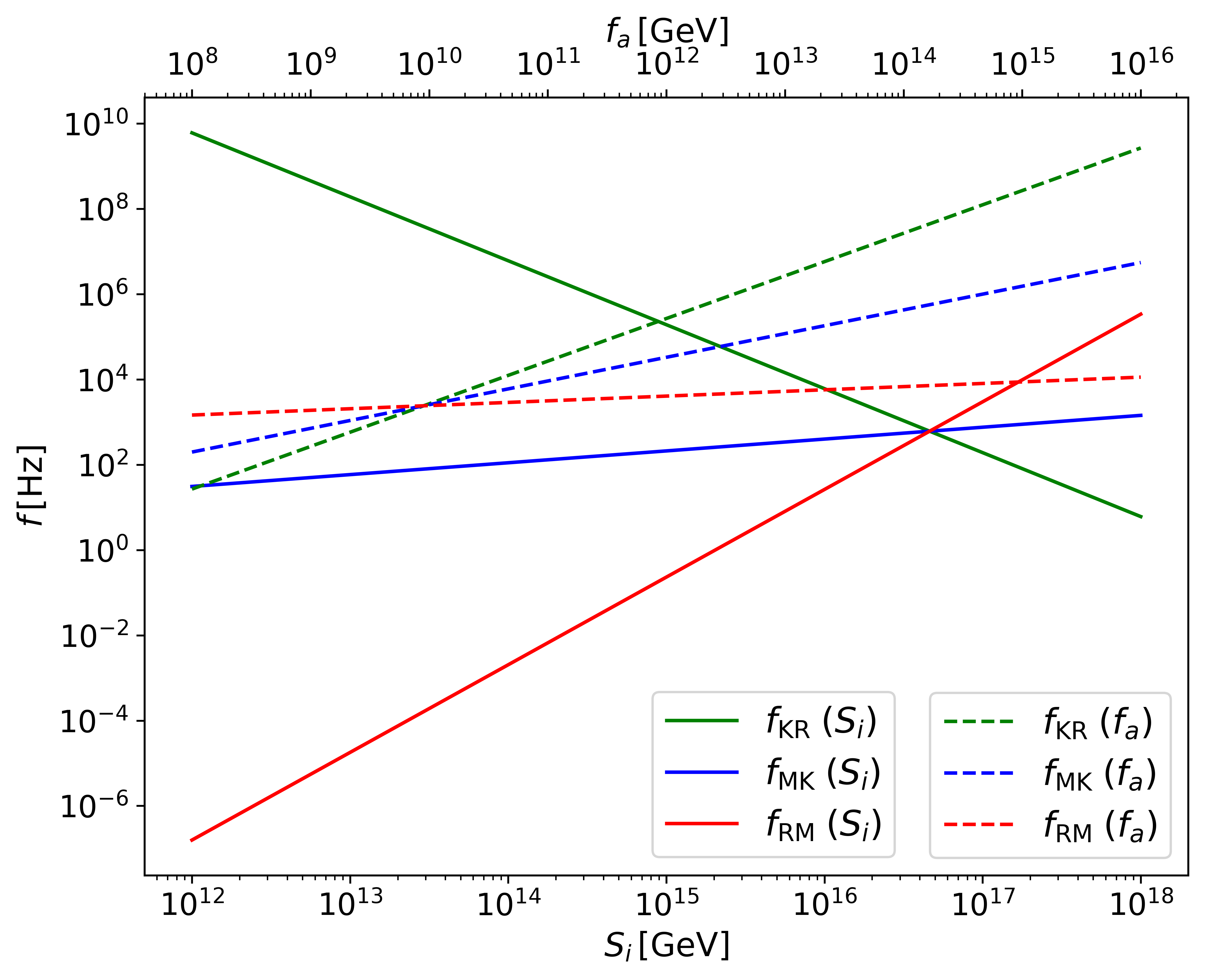}
    \caption{Peak frequency as a function of $S_i$ and $f_a$. We employ $m=4$, $n=1$, $\lambda_n=10^{-10}$, $\delta_n=0$, $\lambda=10^{-10}$, and $\alpha=0.1$. The solid and dashed lines correspond to $f_a=10^{9}~\mathrm{GeV}$ and $S_i=8\times 10^{16}~\mathrm{GeV}$, respectively.}
    \label{fig:peak_frequency_Si_fa}
\end{figure}

\begin{figure}[!htp]
    \centering
    \includegraphics[width=0.4\textwidth]{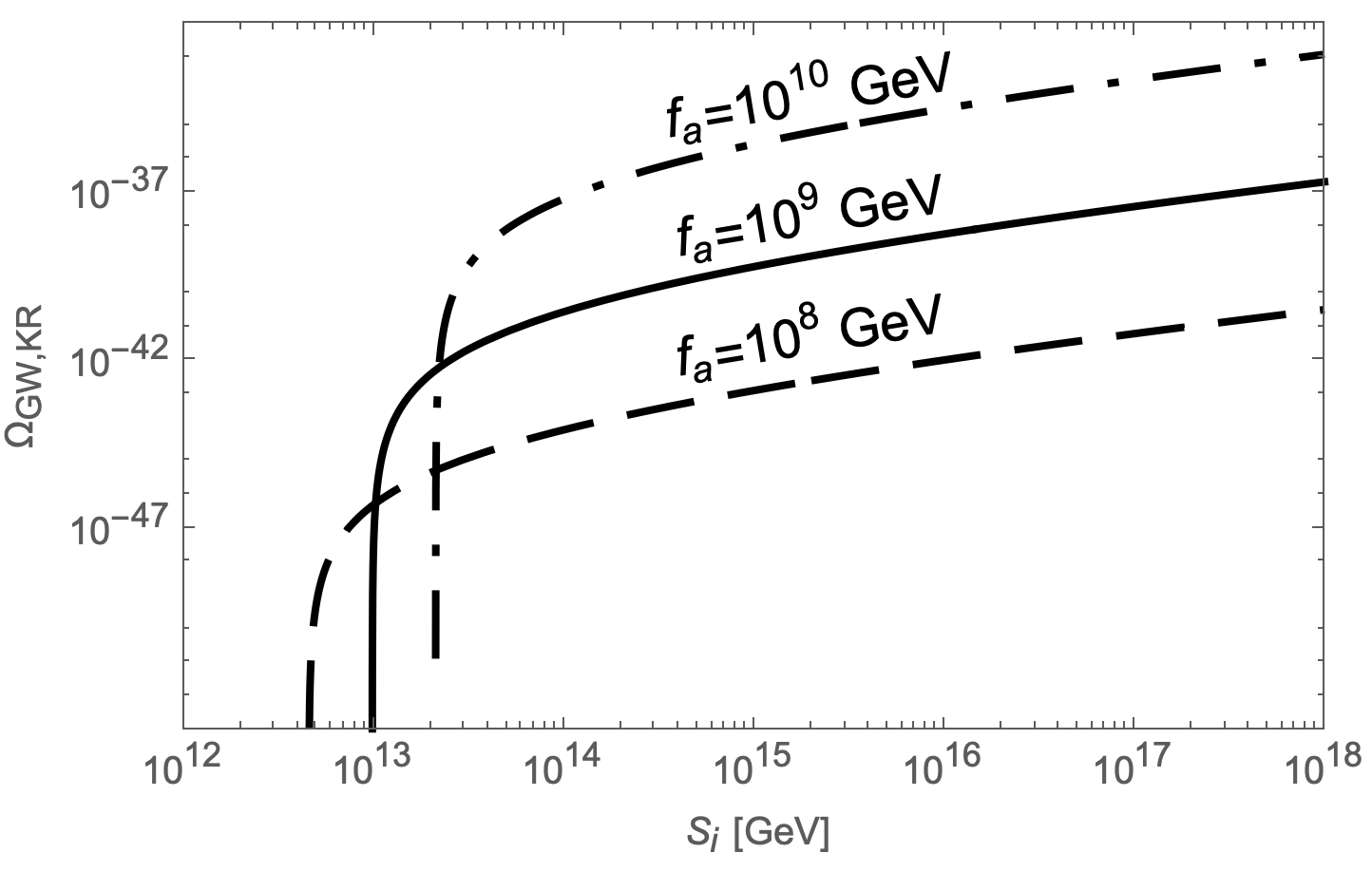}
    \caption{GW peak amplitude as a function of the initial field value. We employ different values of $f_a$, while keeping all other parameters the same as in FIG.~\ref{fig:peak_frequency_Si_fa}.}
    \label{fig:GW_amplitude}
\end{figure}

The spectrum follows an $f^{-1}$ dependence during the kination-dominated era, while the $f^{-1/3}$ behavior in the early matter-dominated epoch arises from the contribution of high-frequency emission modes~\cite{Gouttenoire:2019kij,Cui:2019kkd,Blasi:2020wpy}.
The GW signal is practically undetectable unless significant fine-tuning is invoked. In FIG.~\ref{fig:peak_frequency_Si_fa}, the solid lines illustrate the relation between peak frequency and the initial field value $S_i$. 
The correct order of the cosmological evolution (see FIG.~\ref{fig:kination}) requires $T_\mathrm{RM}>T_\mathrm{MK}>T_\mathrm{KR}$, or equivalently $f_\mathrm{RM} > f_\mathrm{MK} > f_\mathrm{KR}$, leaving only a very narrow viable parameter space. Using Eq.~(\ref{Sec3_D_GW}), the corresponding peak amplitudes of the GW signal are shown in FIG.~\ref{fig:GW_amplitude}, and the signal is exceedingly weak, even for large values of $f_a$. As indicated by Eq.~(\ref{Sec3_D_GW}), one approximately has $\Omega_{\mathrm{GW,KR}} \propto f_a^4$, implying that a significant GW signal would require a large $f_a$. However, Eqs.~(\ref{Sec3_C_TRM}),~(\ref{Sec3_C_TMK}) and~(\ref{Sec3_C_TKR}) show that a large $f_a$ tends to break the temperature hierarchy $T_\mathrm{RM} > T_\mathrm{MK} > T_\mathrm{KR}$. This behavior is also evident from the dashed lines in FIG.~\ref{fig:peak_frequency_Si_fa}, which indicate that the correct cosmological sequence favors smaller values of $f_a$.

In the preceding analysis, we started from the Lagrangian level and fixed a subset of parameters to study the GW signal sourced by global strings, which differs from Ref.~\cite{Co:2021lkc,Gouttenoire:2021jhk}, where the discussion was formulated in terms of intermediate variables. Consequently, we now perform a scan over the Lagrangian parameters to determine whether there exists a region of parameter space that satisfies all of the experimental constraints discussed above. We employ the following ranges

\begin{equation}
\begin{aligned}
    &f_a: 10^8 - 10^{16} \mathrm{GeV}~, \quad S_i: 10^{12} - 10^{18} \mathrm{GeV}~, \\
    &\lambda, \lambda_n: 10^{-30} - 10^{-10}~, \quad \delta_n: 0 - 2\pi~, \\
    &m: 1 - 10~, \quad n: 1 - 10~,\quad \epsilon: 0 - 1~.
    \nonumber
\end{aligned}
\end{equation}

\begin{figure}[!htp]
    \centering
    \includegraphics[width=0.4\textwidth]{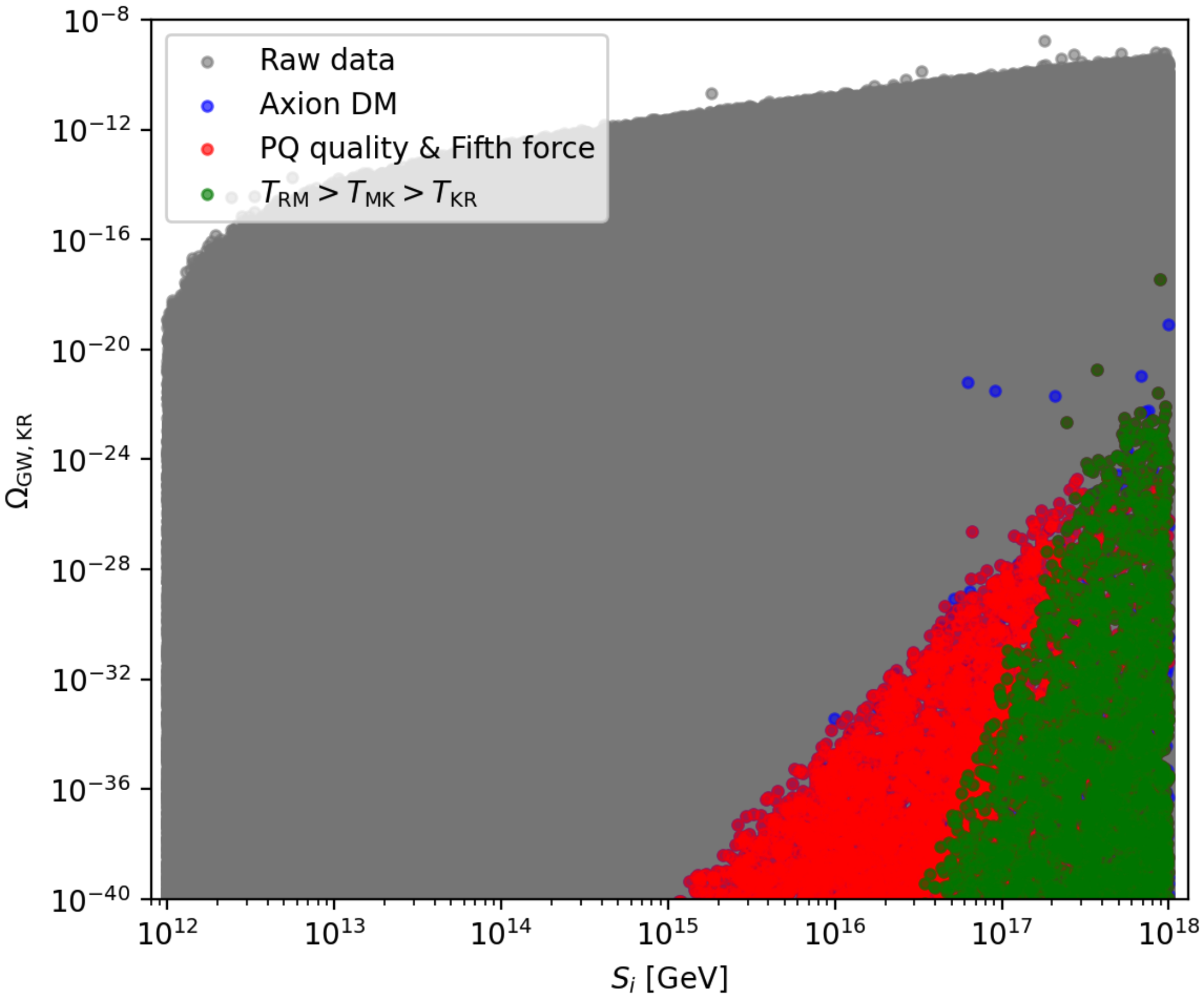}
    \caption{Parameter space satisfying all experimental constraints. The gray points are the raw data. The blue, red, and green points satisfy the saturated relic density of axion DM, PQ quality and fifth-force constrains, and temperature hierarchy, respectively.}
    \label{fig:parameter_space}
\end{figure}

FIG.~\ref{fig:parameter_space} shows the relation between the initial field value $S_i$ and the GW peak amplitude. The grey points represent the raw data. The blue points are a subset of these that satisfy the relic density of axion DM. The red points further impose the constraints from PQ quality and fifth-force bounds. Finally, the green points denote the parameters that additionally satisfy the correct sequence of cosmological evolution, namely $T_\mathrm{RM}>T_\mathrm{MK}>T_\mathrm{KR}$ or equivalently $f_{\rm RM} > f_{\rm MK} > f_{\rm KR}$. As shown in the figure, the resulting GW signal is extremely weak, with $\Omega_{\mathrm{GW},\mathrm{KR}} \lesssim 10^{-20}$,  and thus far beyond the sensitivity of current experiments. 
This is consistent with the discussion before, obtaining a significant GW signal requires a large $f_a$, while such values are incompatible with the correct temperature hierarchy.
In Table.~\ref{table:benchmark}, we present the benchmark points that satisfy all experimental constraints.
\begin{table*}[t]
\small
\centering
\begin{tabular}{|c|c|c|c|c|c|c|c|c|}
\hline
$\lambda$ & $\lambda_n$ & $\delta_n$ & $m$ & $n$ & $\epsilon$ & $f_a$ (GeV) & $S_i$ (GeV) & $\Omega_a h^2$ \\
\hline
$3.7386\times10^{-21}$ &
$5.0335\times10^{-21}$ &
$5.638$ &
$4$ &
$2$ &
$0.9127$ &
$2.4869\times10^{9}$ &
$1.6616\times10^{17}$ &
$0.1205$ \\
\hline
$\overline{\theta}$ & $T_{\rm RM}$ [GeV] & $T_{\rm MK}$ [GeV] & $T_{\rm KR}$ [GeV] & $\Omega_{\mathrm{GW,KR}}$ &
$f_{\rm RM}$ [Hz] & $f_{\rm MK}$ [Hz] & $f_{\rm KR}$ [Hz] & $g_{\rm aN}$\\
\hline
$7.4101\times10^{-35}$ &
$1.6937\times10^{8}$ &
$1.2303\times10^{7}$ &
$3.3160\times10^{6}$ &
$4.5858\times10^{-36}$ &
$20.434$ &
$3.5576$ &
$0.6194$ &
$5.9594\times10^{-46}$ \\
\hline
\end{tabular}
\caption{Benchmark points satisfying all experimental constraints.}
\label{table:benchmark}
\end{table*}
\section{Conclusion}\label{Sec5}

The kinetic misalignment mechanism is a novel production way for axion DM that significantly extends the conventional misalignment scenario. Its central feature is that the axion field is allowed to possess a nonzero initial velocity $(\dot{\theta}\neq 0)$, which substantially delays the onset of oscillations. As a result, the saturated relic density can be achieved for axions with larger masses or smaller decay constants, thereby relaxing the stringent mass or decay constant constraint inherent in the standard misalignment mechanism. Such an initial velocity can naturally arise from explicit PQ symmetry breaking in the early Universe.

Motivated by this, we investigated the axion kinetic misalignment mechanism in the presence of generic PQ-breaking higher-dimensional operators. Starting from a Lagrangian parameters, we studied the coupled dynamics of the axion and the radial mode, which drive an elliptical trajectory in field space. This dynamics naturally introduces an early matter-dominated epoch followed by a kination era into an otherwise radiation-dominated cosmology, thereby modifying the cosmological evolution.
We first analyzed how the initial velocity and the radial oscillations determine the relic density of axion DM. We then examined the associated experimental constraints: the PQ quality bound from the neutron EDM, the implications for axion-mediated fifth-forces, and the requirement that the critical temperatures of the non-standard eras remain compatible with BBN and CMB limits. We also discussed the GW spectrum from global cosmic strings in this modified cosmological history. Although the inserted matter and kination epochs distort the spectrum, their short duration leads to a peak amplitude that well beyond the reach of current experiment sensitivity.
Finally, we performed a parameter space scan over the Lagrangian inputs and identified the region that simultaneously satisfies the DM relic density, PQ quality, fifth-force bounds, BBN and CMB constraints, and the temperature hierarchy $T_\mathrm{RM}>T_\mathrm{MK}>T_\mathrm{KR}$. The allowed parameter space is found to be highly restricted. In particular, enhancing the GW signal would require a large axion decay constant, but such values tend to spoil the required temperature ordering, leaving only extremely suppressed GW amplitudes $\Omega_{\mathrm{GW},\mathrm{KR}} \lesssim 10^{-20}$ within the viable region. We presented benchmark points that satisfy all experimental constraints.

\newpage

\acknowledgments
This research is supported in part by the National Natural Science Foundation of China (NSFC) under Grants Nos. 12322505, 12547101. We also acknowledges Chongqing Talents: Exceptional Young Talents Project No. cstc2024ycjh-bgzxm0020 and Chongqing Natural Science Foundation under Grant No. CSTB2024NSCQ-JQX0022.

\end{document}